\documentclass[11pt]{article}
\usepackage[]{sao1}
\usepackage{graphicx}

\begin{document}

\title{An analysis of the magnetic field geometry and its interaction with the circumstellar environment of HD\,57682 by the MiMeS Collaboration}
\author{J.H. Grunhut\inst{1} \and  G.A. Wade\inst{1} \and W.L.F. Marcolino\inst{2} \and  V. Petit\inst{3} \and  H.F. Henrichs\inst{4} \and and the MiMeS Collaboration}
\institute{Department of Physics, Royal Military College of Canada, Kingston, Ontario, Canada\and LAM-UMR, CNRS \& Univ. de Provence, Marseille, France \and Department of Geology and Astronomy, West Chester University, West Chester, PA, USA \and Astronomical Institute ``Anton Pannekoek", University of Amsterdam, Amsterdam, Netherlands}

\maketitle 

\begin{abstract}
We will review our recent analysis of the magnetic properties of the O9IV star HD\,57682, using spectropolarimetric observations obtained with ESPaDOnS at the Canada-France-Hawaii telescope within the context of the Magnetism in Massive Stars (MiMeS) Large Program. We discuss our most recent determination of the rotational period from longitudinal magnetic field measurements and H$\alpha$ variability - the latter obtained from over a decade's worth of professional and amateur spectroscopic observations. Lastly, we report on our investigation of the magnetic field geometry and the effects of the field on the circumstellar environment.
\keywords{instrumentation: polarimeters, techniques: spectroscopic, stars: magnetic fields, stars: rotation, stars: individual (HD 57682)}
\end{abstract}

\section{Introduction}
\begin{figure} 
\centering
\includegraphics[width=2.5in]{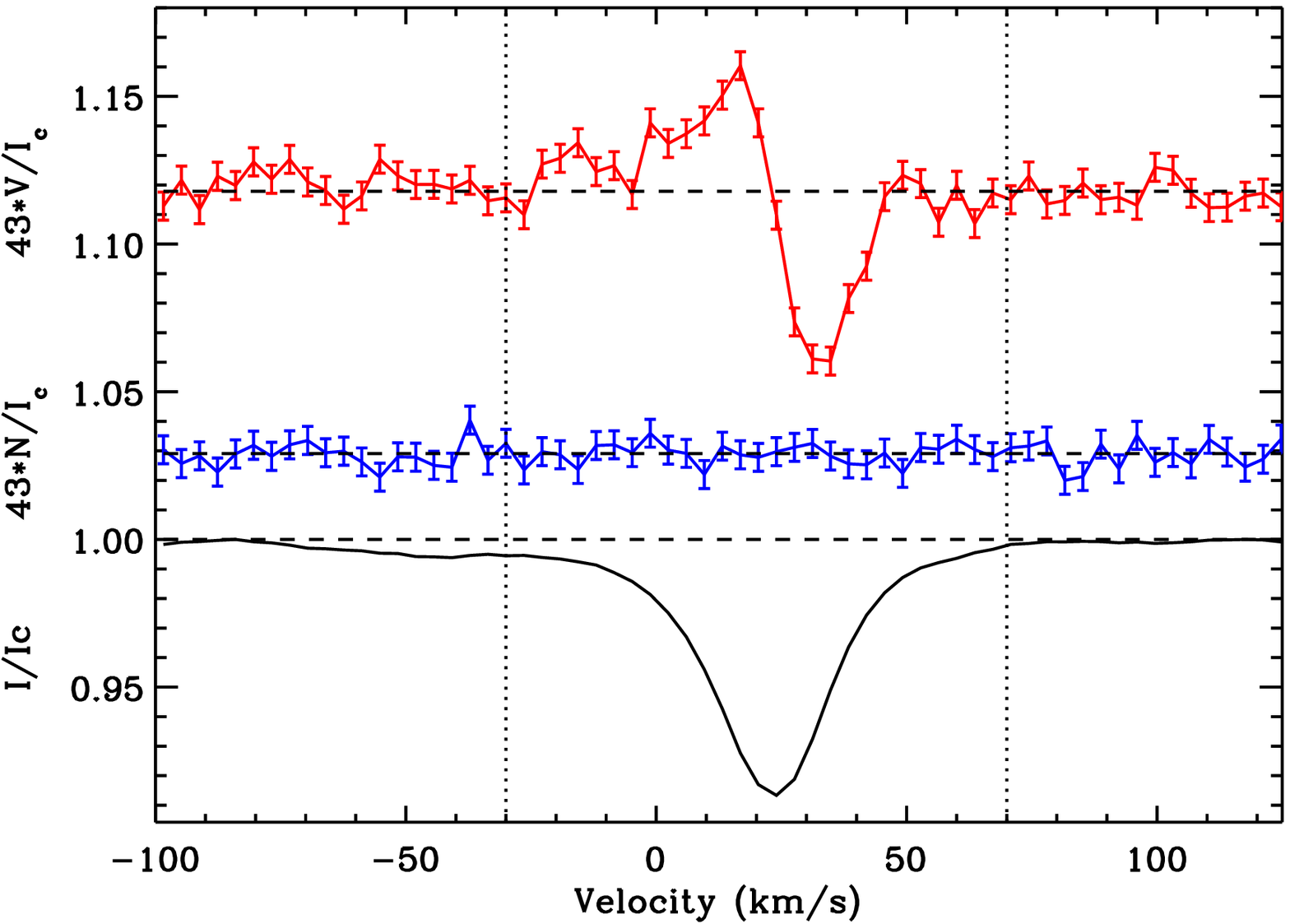}
\includegraphics[width=3.25in]{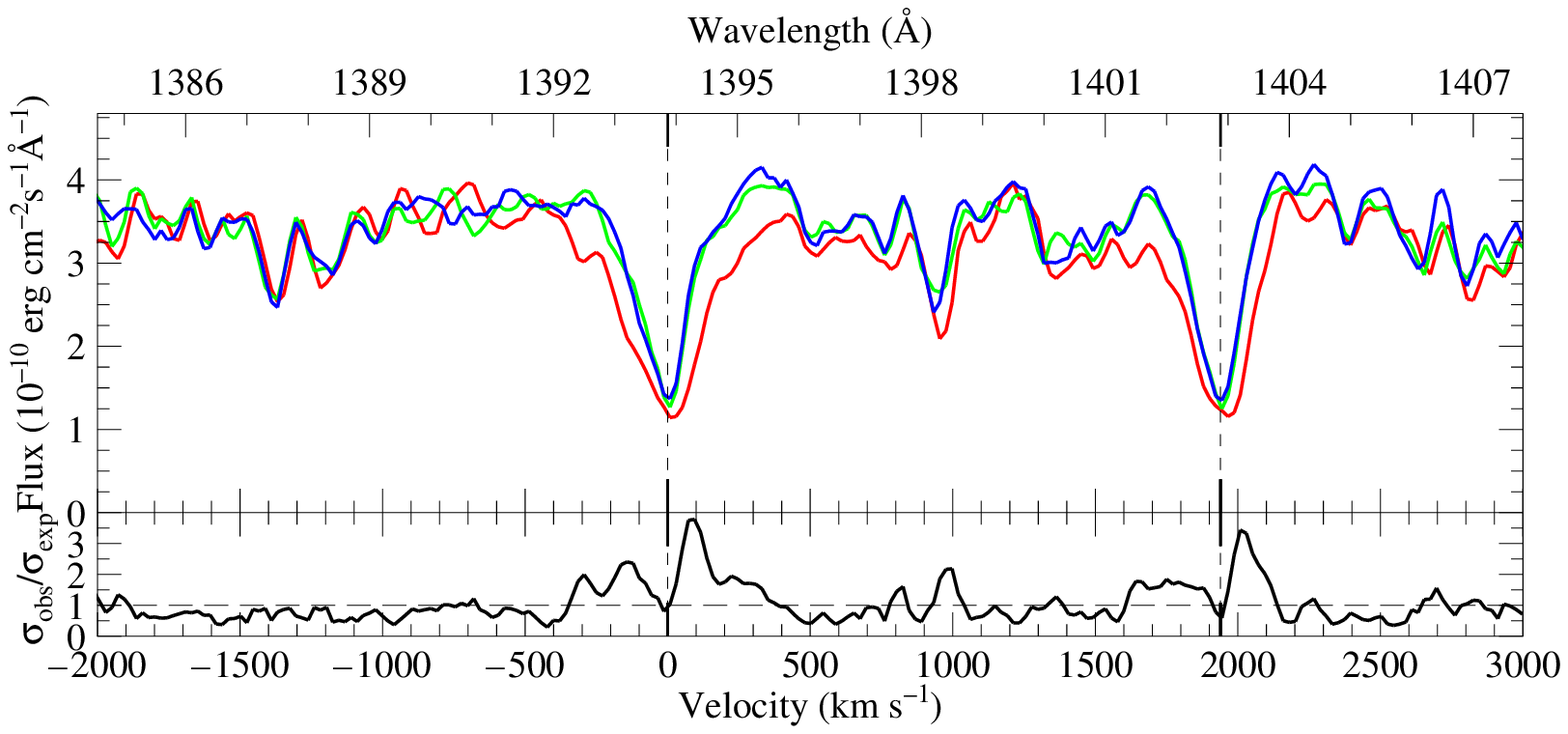}
\caption{{\bf Left:} Observed mean LSD Stokes $V$ (top), diagnostic null (middle), and unpolarized Stokes $I$ (bottom) profiles of HD\,57682 from 2008-12-06. {\bf Right:} Overplot of 3 $IUE$ UV spectra of the Si\,{\sc IV} line profiles (top). The significance of the variability is displayed at the bottom.}
\label{fig1}
\end{figure}

The presence of strong, globally-organized magnetic fields in hot, massive stars is rare. To date, only a handful of massive O-type stars are known to host magnetic fields. In 2009, Grunhut et al.  reported the discovery of a strong magnetic field in the weak-wind O9IV star HD\,57682 from the presence of Zeeman signatures in mean Least-Squares Deconvolved (LSD) Stokes~$V$ profiles (see Fig.~\ref{fig1}). Grunhut et al. also used $IUE$ ultraviolet and ESPaDOnS optical spectroscopy to determine the following physical parameters using CMFGEN: $T_{\rm eff}=34.5$\,kK, $\log(g)=4.0\pm0.2$, $R=7.0^{+2.4}_{-1.8}$\,$R_{\odot}$, $M=17^{+19}_{-9}$\,$M_{\odot}$, and $\log({\dot{M}})=-8.85\pm0.5$\,$M_{\odot}$\,yr$^{-1}$. Of particular interest, we highlight the low mass-loss rate derived from UV wind diagnostic lines, which show variability characteristic of other magnetic OB stars (e.g. Schnerr et al. 2008), as shown in Fig.~\ref{fig1}.

With only 7 observations of this star at our disposal at that time we employed a Bayesian inference method to determine the best-fitting dipole parameters to characterize the magnetic field (Petit et al. in prep). We concluded that the dipolar field of HD\,57682 is characterized by a polar strength of $\sim$1.7 kG, and a magnetic axis aligned within 10 to 50$^{\circ}$ of the rotation axis, depending on the inclination of the rotation axis to our line of sight.

Since our original report of this star in 2009, we have obtained an additional 10 high-resolution spectropolarimetric observations with the ESPaDOnS instrument at CFHT.

\section{Temporal Variability}
\begin{figure}
\centering
\includegraphics[width=6.25in]{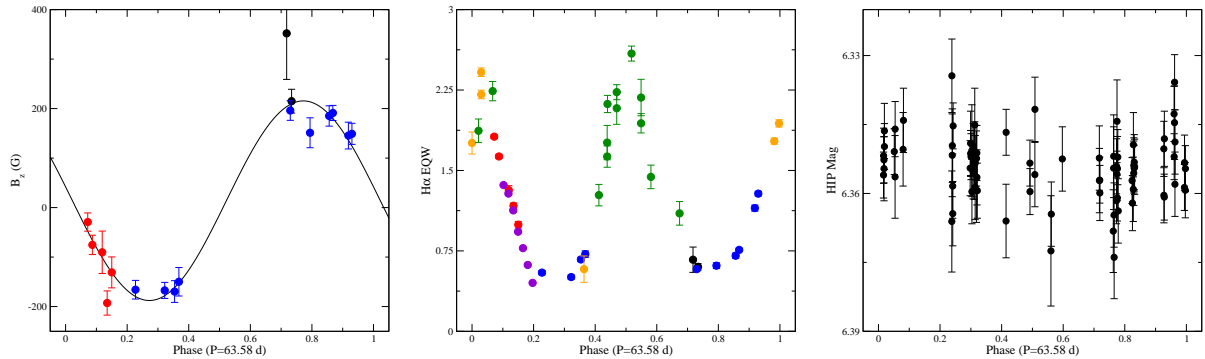}
\caption{Phased longitudinal magnetic field measurements (left), H$\alpha$ equivalent width variation (middle), and Hipparcos photometry (right), for HD\,57682. Different colours indicate different epochs of observations.}
\label{sidebyside}
\end{figure}

Both the longitudinal magnetic field and H$\alpha$ equivalent width of HD\,57682 are strongly variable. In addition to our 17 ESPaDOnS observations, we've also utilized H$\alpha$ observations from amateur spectroscopy from the BeSS database, as well as archival ESO UVES and FEROS observations dating back over a decade. A period search of these data resulted in a period of $\sim$31\,d, consistent with the maximum original rotational period estimated by Grunhut et al. (2009). However, the magnetic data could not be reasonably phased with this period. Ultimately, adopting a period of 63.58\,d (twice the period obtained from the H$\alpha$ data) resulted in a coherent phasing of all the data at our disposal, as shown in Fig.~\ref{sidebyside}.This period is inconsistent with the fundamental parameters derived by Grunhut et al., likely indicating that the $v\sin i$ is too high by nearly a factor of two.

The longitudinal magnetic field appears to vary sinusoidally, consistent with a magnetic field dominated by a strong dipolar component. The H$\alpha$ equivalent width shows a double-wave pattern with peak emission occurring at the magnetic crossover phases (i.e. when the longitudinal field is null).

The photometric light curve from Hipparcos shows no apparent variability. This likely indicates that the column density of the magnetically confined plasma is relatively low at eclipse phases.

\section{Magnetic Geometry and Circumstellar Environment}
We are capable of fitting a dipole model (characterized by the magnetic field strength at the poles ($B_d$) and the angle of obliquity of the magnetic axis relative to the rotation axis ($\beta$)) to the longitudinal component of the magnetic field, measured from the mean LSD Stokes $V$ and $I$ profiles. However, since we do not have any constraints on the inclination angle of the rotation axis, we can only estimate that $B_d=1-3$\,kG and $\beta=60-90^{\circ}$.

\begin{figure}
\centering
\includegraphics[width=2.8in]{grunhut_hd57682_fig3.eps}
\includegraphics[width=2.68in]{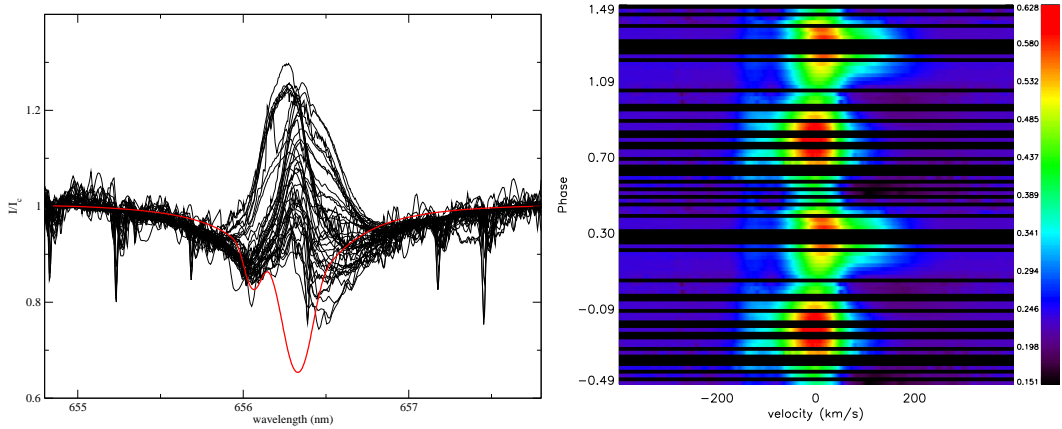}
\caption{{\bf Left:} H$\alpha$ (black) profiles for different nights compared to a LTE model profile (red). {\bf Right:} Phased H$\alpha$ residual variations relative to an LTE model. Note the variability is likely due to a magnetically confined wind.}
\label{chisq}
\end{figure}

In Fig.~\ref{chisq} we show the residual variations of H$\alpha$ phased with the adopted rotational period. We conclude based on the characteristics of this dynamic spectrum that the magnetic field is exerting strong confinement (confinement parameter $\eta*\sim10^3-10^5$; ud-Doula \& Owocki 2002) on the weak wind of HD\,57682, resulting in the observed H$\alpha$ variability. However, the slow rotation is likely unable to centrifugally support a stable magnetosphere. Therefore, the plasma that is present likely has a relatively short residence time in the magnetosphere, and must therefore be continually replenished.

\end{document}